\documentclass[12pt]{iopart}
\usepackage[dvips]{graphicx}

\begin{document}

\title{Roughening and pinning of interface cracks in shear delamination of thin films}

\textbf{}

\textbf{}

\author{M. Zaiser$^1$, P. Moretti$^{1,2}$, A. Konstantinidis$^3$, E.C. Aifantis$^{3,4}$}

\textbf{}

\textbf{}

\address{$^1$The University of Edinburgh, Center for Materials Science and Engineering, The King's Buildings, Sanderson Building, Edinburgh EH93JL, UK}

\address{$^2$Departament de Fisica Fonamental, Facultat de Fisica, Universitat de Barcelona, Marti i Franques 1, E-08028 Barcelona, Spain}

\address{$^3$Laboratory of Mechanics and Materials, Aristotle University of Thessaloniki, 54124 Thessaloniki, Greece}

\address{$^4$Center for the Mechanics of Material Instabilities and Manufacturing Processes (MMIMP), Michigan Tech., Houghton, USA}

\eads{\mailto{M.Zaiser@ed.ac.uk}, \mailto{paolo.moretti@ub.edu},
  \mailto{akonsta@gen.auth.gr }, \mailto{mom@mom.gen.auth.gr}}
\begin{abstract}

We investigate the roughening of shear cracks running along the interface between a thin film and a rigid substrate. We demonstrate that short-range correlated fluctuations of the interface strength lead to self-affine roughening of the crack front as the driving force (the applied shear stress/stress intensity factor) increases towards a critical value. We investigate the  disorder-induced perturbations of the crack displacement field and crack energy, and use the results to determine the crack pinning force and to assess the shape of the critical crack. The analytical arguments are validated by comparison with simulations of interface cracking. 
\end{abstract}

\section{Introduction}

Strength of materials is governed by the properties of defects. In particular, in order to understand materials failure, one needs to study the nucleation of new and the propagation of existing cracks. While many investigations (for an overview, see  \cite{alava06a}) have focused on cracks in bulk materials, in the present work we consider near-surface cracks running along a weak interface which separates a thin elastic film from a rigid substrate and acts as a prescribed fracture surface. Failure is induced by shear tractions that are applied to the free surface of the film and induce shear stresses in the film and on the interface. The considered geometry applies to a variety of materials problems, such as shear failure and/or abrasive wear of coatings and shear-induced delamination of thin films. One may also consider geophysical applications where the load acting on the weak interface originates from gravitational body forces, such as in the initiation of snow slab avalanches by failure of weak interfaces in snow stratifications \cite{zaiser04a,fyffe04a,heierli06,heierli08}. 

While in a companion paper \cite{zaiser09} we have addressed the problem of crack nucleation, in the present study we focus on the behavior of pre-existing interface cracks. We assume a disordered interface which exhibits a random pattern of weak and strong regions. Hence, the shape of an advancing crack becomes distorted as the crack advances across weak regions and gets arrested at regions of elevated strength. At the same time, elastic forces resist the distortion and try to keep the crack front straight. The interplay between crack front elasticity and disorder gives rise to the mutually related phenomena of crack front roughening and crack pinning. 

The roughening of bulk cracks has been investigated extensively in view of explaining the emergence of self-affine fracture surfaces (see e.g. \cite{bouchaud03}, \cite{alava06b}). The long-range elastic self-interaction of the two-dimensionally perturbed front of a bulk crack was analyzed by Gao and Rice \cite{gao89} and the implications on quasi-static crack propagation in disordered materials were discussed by Ramananthan et. al. \cite{fisher97}. In this work, cracks were envisaged as one-dimensional elastic manifolds with long-range self-interactions, which are driven by an external force (the applied stress intensity factor) through a 'landscape' of short-range correlated disorder. From this point of view, the in-plane roughening of the crack front is expected to be governed by the same roughness, correlation length and dynamic exponents which characterize the depinning of a contact line. These exponents were evaluated by Ertas and Kardar using a one-loop renormalization scheme \cite{ertas94}. A two-loop renormalization calculation of the same exponents was done by Chauve et. al. \cite{chauve01,ledoussal02}. While results of the one-loop calculation yield roughness exponents that are generally too low when compared with experimental data, the two-loop correction leads to an in-plane roughness exponent of about $\zeta = 0.5$ which is in line with available experimental observations.

From an experimental point of view, the out-of-plane roughness of bulk cracks (i.e., the roughness of the fracture surface) has been studied in many papers, whereas comparatively few studies have been devoted to investigating in-plane roughness, i.e. the shape of the crack front in the direction of crack propagation. The obvious reason is that observing the crack front requires tricky {\em in situ} methods, whereas the out-of-plane roughness of the crack transfers to the fracture surface which can be easily accessed after the specimen has failed. In-plane roughening of crack fronts was studied by Delaplace et. al. \cite{schmittbuhl99} for cracks running along the interface between two PMMA plates. These authors report typical roughness exponents of $\zeta \approx 0.5-0.6$. The burst-like dynamics of crack propagation was studied by Vellinga et. al. \cite{vellinga06} who used image correlation methods to monitor the propagation of a crack along the interface between a glass subtrate and a polymer film. 

The above mentioned theoretical investigations refer to bulk cracks. In thin film geometries, on the other hand, the range of elastic interactions is limited by the film thickness and this modifies the energy functional for shear cracks \cite{louchet02} as well as for tensile cracks \cite{astrom00} and anticracks \cite{heierli06,heierli08}. Crack front roughening in thin film geometries was investigated by numerical simulation of a tensile crack running between two elastic sheets \cite{astrom00}, and by analytical arguments and numerical simulation of a shear crack underneath a thin elastic substrate \cite{zaiser04b}. In the present paper we further develop this investigation and show that the roughening of crack fronts in thin film geometries is governed by a complicated interplay of length scales. In Section 2 we determine the elastic energy functional of a thin film with a general interfacial displacement field and use this to evaluate displacement fields and elastic energies of straight and perturbed mode-II cracks. We then use in Section 3 standard scaling arguments from pinning theory to determine the pinning length above which the crack roughens, and evaluate the associated increase in the critical crack driving force. Furthermore we show that the presence of the opposite crack front leads to emergence of a second critical length above which the crack is flat again. The scaling arguments are validated by comparison with simulation results (Section 4).

Most investigations of crack front roughening envisage the dynamics of the crack front as a function of the driving force (the stress intensity factor). However, the stress intensity factor is itself an increasing function of the crack length and the acting stress. In experimental investigations the driving force can be controlled in such a manner that stable crack propagation takes place \cite{schmittbuhl99, vellinga06}. In practice, however, systems are often subject to fixed loads. In this case, the increase of driving force with increasing crack length leads to catastrophic failure by unstable crack propagation once a critical stress threshold is reached. In our discussion section we assess the influence of disorder on the failure stress of a cracked interface under conditions of load control. We demonstrate that the advance of the crack front during roughening leads to complex behavior where increasing the disorder may either result in weakening by disorder-induced crack advance or in strengthening by crack pinning, while at very high disorder a change in failure mode is observed.  

\section{Formulation of the elastic problem}

\subsection{General considerations}

We consider an elastic film of thickness $D$ tethered to a rigid substrate. The interface with the substrate is modeled as a cohesive layer in the plane $z=0$. The response of the interface to shear loads is characterized by the scalar stress-displacement relationship $\sigma _{xz} (x,y,z=0)=\tau (u(x,y))$ where $\sigma _{xz} $ is the shear stress at the interface, $\tau $ is the interfacial shear strength, and  $u(x,y)=w_{x} (x,y,z=0)$ is the shear displacement across the interface. The maximum stress that can be supported by the interface is denoted as $\tau _{M} $, and the specific fracture energy per unit interface area is given by the integral
\begin{equation}
W_{\rm f} =\int \tau (u){\rm d}u=:\tau _{{\rm M}} u_{0} \;,
\label{wf}
\end{equation}
where $u_0$ denotes the characteristic displacement-to-failure. Structural disorder of the interface is modelled in terms of 
short-range correlated fluctuations of the fracture energy $W_{\rm f} = W_{\rm f}(x,y)$. The stochastic process $W_{\rm f}(x,y)$ is assumed to have finite first and second moments $\langle W_{\rm f} \rangle$ and $\langle W_{\rm f}^2 \rangle$, and finite correlation length $\xi$.  

The film is loaded by spatially homogeneous tractions applied to its free surface at $z=D$, giving rise to a space-independent 'external' shear stresss $\sigma _{xz}^{{\rm EXT}}$. This external stress superimposes on the internal stresses that are associated with the interfacial displacement field $u(x,y)$. To evaluate these stresses we start from the elastic energy functional associated with a generic displacement vector field $\bi{w}(\bi{r})$. This functional is given by
\begin{equation}
H({\bf w})=\frac{1}{2}\int \left[ \lambda \left(\sum_i w_{ii}\right)^2 + 2 \mu \left(\sum_{ij} \left(w_{ij}\right)^2\right) \right] d^3r,
\label{Hw}
\end{equation}
where the $w_{ij}$ are the components of the strain tensor, $w_{ij}=(\partial_i w_j + \partial_j w_i)/2$, and indices assume the values $\{i,j\}\in \{x,y,z\}$. The parameter $\lambda$ can be expressed in terms of the shear modulus $\mu$ and Poisson's ratio $\nu$ as $\lambda=2\mu\nu/(1-2\nu)$. Energy minimization gives the equilibrium equation
\begin{equation}
\nabla ^{2} {\bi{w}}+\frac{1}{1-2\nu } {\rm grad} \left({\rm div}\, {\bi{w}}\right)=0\;.
\label{equil}
\end{equation}
We have to solve this equation for the boundary conditions $w_{x} (x,y,z=0)=u(x,y),\; w_{y} =w_{z} =0$. In Fourier space the solution is given by
\begin{eqnarray}
&& \bi{w}=\frac{1}{(2\pi)^2}
\int \left(\begin{array}{r} 1 - \eta z
\frac{k_x^2}{(k_x^2+k_y^2)^{1/2}}\\
-\eta z\frac{k_xk_y}{(k_x^2+k_y^2)^{1/2}}\\
-i\eta z \,k_x\end{array}\right)\nonumber\\
&& \times 
\exp\left[i(k_xx+k_yy)-(k_x^2+k_y^2)^{1/2}z\right]\,u(k_x,k_y)\, {\rm d}^2 k\;,
\end{eqnarray}
where $u(k_x,k_y)$ is the Fourier transform of $u(x,y)$ and $\eta = 1/(3-4\nu)$. In the present work we are interested in the behavior of interface cracks underneath thin films. We therefore assume that the film thickness $D$ is much smaller than the characteristic length of variations in the displacement field, such that  $\left|kz\right|<<1$ for all $z \le D$. The displacement field is then approximately given by
\begin{equation}
\bi{w}=\frac{1}{(2\pi)^2}
\int \left(\begin{array}{r} 1 \\
0\\
0 \end{array}\right)
\exp\left[i(k_xx+k_yy)-(k_x^2+k_y^2)^{1/2}z\right]\,u(k_x,k_y)\,{\rm d}^2 k\;.
\end{equation}
Inserting this into Eqn. (\ref{Hw}) and using again that $kD \ll 1$ gives
\begin{equation}
H(u)=\frac{\mu D}{(2\pi)^2} \int
\left[\left(\frac{1+\alpha}{2}\right)k_x^2+k_y^2\right]
u(k_x,k_y)u(-k_x,-k_y)\; {\rm d}^2 k\;,
\end{equation}
where  $\alpha=(2-2\nu)/(1-2\nu)$. Reverting to spatial coordinates, we find the energy functional
\begin{equation}\label{slopeH2}
H(u)=\mu D \int dx\int dy
\left[\left(\frac{1+\alpha}{2}\right) (\partial_x u)^2 +
(\partial_y u)^2 \right]\;.
\label{Hu}
\end{equation}
The total energy of the system is obtained by adding to the elastic energy $H$ the work done by the shear stresses at the interface: 
\begin{equation}
G(u)=-\int {\rm d}x\left[\int _{0}^{u(x)}\left(\sigma _{xz}^{{\rm EXT}} -\tau (u)\right)\; {\rm d}u \right] \;.
\label{Gu}
\end{equation}
Minimizing the total energy functional $E(u) = G(u) + H(u)$ leads to the equilibrium condition 
\begin{equation}
I_x\frac{\partial ^{2} u}{\partial x^{2} }+ I_y\frac{\partial^{2} u}{\partial y^{2} } +\sigma _{xz}^{{\rm EXT}} -\tau (u) = 0\;,
\label{equil_u}
\end{equation}
where the mode-II and mode-III interaction constants $I_x$ and $I_y$ are given by $I_x = D\mu(1+\alpha)$ and $I_y = 2 D\mu$.  

\subsection{Displacement field of a mode-II shear crack and crack energy functional for perturbations of the crack front}

To analyse the behavior of a crack in a disordered medium, we first evaluate the properties of a straight mode-II crack and then study crack front roughening in terms of perturbations of this reference crack. 

We consider a crack of width $2l$ with crack fronts located at $x=\pm l$. In the brittle limit $\tau(u)=  W_{\rm f} \delta(u)$ ($\delta(u)$ is Dirac's delta function) the corresponding displacement field $u_0$ is given by
\begin{equation}
u_0(x)=\left\{
\begin{array}{l l }\displaystyle\frac{\left(l^{2} -x^{2} \right)\sigma _{xz}^{{\rm EXT}} }{2I_x}\;,& x \in [-l,l]\;,\\
0 & {\rm else}\;.
\end{array} 
\right.
\label{ux}
\end{equation}
Inserting the displacement field, Eqn. (\ref{ux}), into the energy functional, Eq. (\ref{Hu}), gives the elastic energy of the reference crack:
\begin{equation}
H(l) = -\frac{l^{3}}{3I_x} \left(\sigma _{xz}^{{\rm EXT}} \right)^{2}\;.
\label{Hl}
\end{equation}
The derivative of the elastic energy with respect to $l$ is the crack driving force per unit length (the stress intensity factor). Since the crack has two fronts, the driving force for each front is
\begin{equation}
f(l) = \frac{(l \sigma _{xz}^{{\rm EXT}})^{2}}{2I_x}\;.
\label{Fl}
\end{equation}
This driving force is resisted by the fracture energy of the interface. For the reference crack with 
fracture energy $\langle W_{\rm f} \rangle$, equating $f$ and $\langle W_{\rm f} \rangle$ defines a critical crack length
\begin{equation}
l_{\rm c} = \sqrt{2 \langle W_{\rm f}\rangle I_x}/\sigma_{xz}^{\rm EXT}\;.
\end{equation}

We now consider situations where the fracture energy exhibits random spatial fluctuations. In this case, parts of the crack front may advance across regions of reduced fracture energy and then get trapped at locations of elevated fracture energy. Thus, the crack may advance and roughen before it gets critical. To study the energetics of crack front roughening, we envisage a generic perturbation $\delta l(y)$ of the straight crack front at $x=l$. This corresponds to a perturbation $\delta u(x,y)$ of the displacement field which along the front of the unperturbed crack must fulfil the boundary condition 
\begin{equation}
\delta u(x=l,y) = \left(\left.\frac{\partial u_0}{\partial x}\right|_{x=l}\right) \delta l(y)
= \left(\left.\frac{\partial u_0}{\partial x}\right|_{x=l}\right) \int  \exp(i k_y y)\delta l(k_y) \frac{{\rm d}k_y}{2\pi} \;,
\label{bc}
\end{equation}
where $u_0$ is given by Eqn. (\ref{ux}). Over the width of the crack, the displacement field fulfils the partial differential equation 
\begin{equation}
I_x\frac{\partial ^{2} u}{\partial x^{2} }+ I_y\frac{\partial^{2} u}{\partial y^{2} } = 0\;,
\label{equil_u_crack}
\end{equation}
The solution which satisfies the boundary condition (\ref{bc}) is given by
\begin{equation}
u(x,y)=u_0(x)+ \left(\left.\frac{\partial u_0}{\partial x}\right|_{x=l}\right)  \int \exp\left[ik_y y+\sqrt{\frac{I_{y}}{I_{x}}}
k_y(l-x)\right]\delta l(k_y) \frac{{\rm d} k_y}{2\pi}\;.
\end{equation}
By inserting this solution into the elastic energy functional, Eqn. (\ref{Hu}), we can evaluate the energy cost associated with the perturbation. In Fourier space the elastic energy variation associated with the crack front perturbation $\delta l(k_y)$ is given by
\begin{eqnarray}
\delta H &\approx&
\frac{1}{2}\left(\left.\frac{\partial u_0}{\partial x}\right|_{l}\right)^2 \int\sqrt{I_{x}I_{y}}|k_y| |\delta l(k_y)|^2 \frac{{\rm d} k_y}{2 \pi}\nonumber\\
&=& (l \sigma_{xz}^{\rm EXT})^2 \sqrt{\frac{I_y}{4I_x^3}} \int |k_y| |\delta l(k_y)|^2 \frac{{\rm d} k_y}{2 \pi} \;,
\label{cracfront}
\end{eqnarray}
that is, the front of a perturbed interface crack behaves like an elastic manifold with long-range elasticity. Similar behavior can be found for bulk cracks or contact lines \cite{gao89,ertas94}. 

In deriving Eqn. (\ref{cracfront}) we have assumed that $k_y l \gg 1$ such that terms of order $\exp[- 2 k_y l]$ can be neglected. This implies that our treatment applies only to fluctuations with wavelengths that are short in comparison with the crack width. In the opposite case $k_y l \ll 1$, the boundary conditions at the opposite crack front ($x=-l$) need to be considered as well. Assuming that front unperturbed, the crack energy change can be obtained by simply inserting the modified crack length, $l \to l + \delta l(y)$, into Eqn. (\ref{Hl}) and integrating over $y$. For perturbations of zero average this gives in Fourier space
\begin{equation}
\delta H \approx
\frac{l}{2I_x} \left(\sigma _{xz}^{{\rm EXT}} \right)^{2}\int |\delta l(k_y)|^2 \frac{{\rm d} k_y}{2 \pi}\;.
\label{cracfront1}
\end{equation}
This implies that crack elasticity on scales above the crack width becomes of infinite range: the interaction kernel is independent of the wavevector. 

\section{Crack pinning and crack roughening}

We now study the changes in crack geometry caused by the disorder. The characteristic fracture energy fluctuation experienced by a smooth crack front segment of length $\lambda$ is of the order of 
\begin{equation}
\delta W_{\rm f} =  \left(\frac{\xi}{\lambda}\right)^{1/2} \langle W_{\rm f} \rangle C_{\rm V}\;,
\label{cracfluct}
\end{equation}
where $C_{\rm V}$ denotes the coefficient of variation of the fracture energy distribution, $C_{\rm V}^2 = \langle W_{\rm f}^2 \rangle/\langle W_{\rm f}\rangle^2$. The correlation length of the fracture energy fluctuations perpendicular to the crack is given by $\xi$. Crack front roughening on scale $\lambda$ occurs if the energy gain by adjusting to the fluctuating energy profile is larger than the energy that has to be expended to change the crack shape. Combining Eqs. (\ref{cracfluct}) and (\ref{cracfront}), and considering $\delta l$ in Eq. (\ref{cracfront}) as a perturbation of amplitude $\xi$ and characteristic wavelength $\lambda$, we find the condition
\begin{equation}
\delta H(\xi,\lambda) \approx (l \sigma_{xz}^{\rm EXT})^2 \sqrt{\frac{I_y}{4I_x^3}} \xi^2\le 
\xi \lambda \delta W_{\rm f}(\lambda) =  (\xi^3 \lambda)^{1/2} \langle W_{\rm f} \rangle C_{\rm V}\;.
\label{cracfluctnet}
\end{equation}
Hence, roughening occurs above a critical scale 
\begin{equation}
\lambda_{\rm c} \approx \frac{(l \sigma_{xz}^{\rm EXT})^4 I_y \xi}{4 I_x^3 \langle W_{\rm f} \rangle^2 C_{\rm V}^2} 
= \frac{I_y}{I_x}\left(\frac{l}{l_c}\right)^4 \frac{\xi}{C_{\rm V}^2}.
\label{lambdac}
\end{equation}
Above this scale, the crack adjusts to the disordered pinning 'landscape'. This leads to an incresase in the critical driving force needed for initiating supercritical crack propagation or, equivalently, of the effective fracture energy. The magnitude of the increase can be estimated by averaging the fluctuations of the specific fracture energy over the length $\lambda_{\rm c}$ as evaluated for the critical reference crack ($l = l_{\rm c}$): 
\begin{equation}
\Delta f_{\rm c} \approx \left(\frac{\xi}{\lambda_{\rm c}(l_{\rm c})}\right)^{1/2} \langle W_{\rm f} \rangle C_{\rm V}
\approx \left(\frac{I_x}{I_y}\right)^{1/2} \langle W_{\rm f} \rangle C_{\rm V}^2 \;.
\label{cracpin2}
\end{equation}
At a given applied stress, this increase in the critical driving force implies that the critical crack length $l_{\rm c}^*$ in the presence of disorder will somewhat exceed the critical crack length $l_{\rm c}$ in the homogeneous reference system. However, as will be shown below, this increase is in general small. 

Above the scale $\lambda_{\rm c}$, the crack front develops irregular fluctuations. In this regime, the depinning theory for manifolds with long-range elasticity predicts the crack shape to be self-affine with roughness exponent $\zeta \approx 0.47$ \cite{chauve01}. According to this theory, self-affine scaling is expected to extend up to a correlation length which diverges as the crack driving force approaches the critical value where the crack depins: At this point, the crack is expected to be self affine on all scales above $\lambda_{\rm c}$. However, this is not true here, since the nature of the elastic energy functional imposes an intrinsic limit on the extension of the self-affine scaling regime. As soon as we consider scales $\lambda$ that are larger than the crack length $l$, Eqn. (\ref{cracfront}) needs to be replaced by Eq. (\ref{cracfront1}) for calculating the elastic energy of a bulge. Repeating the steps in Eqs. (\ref{cracfluct}) to (\ref{lambdac}), we find that roughening is restricted to scales below 
\begin{equation}
\lambda_{\rm c}^* \approx \frac{l^2}{\lambda_{\rm c}}\;.
\label{lambdastar}
\end{equation}
Above this scale, the elastic interactions again suppress the roughening. Depending on the magnitude of the disorder, this creates two possible scenarios (we assume that $I_x/I_y$ and $l/l_{\rm c}$ are of the order of one):  
\begin{itemize}
\item For $\xi/l > C_{\rm V}^2$ (weak disorder): $\lambda_{\rm c} > l > \lambda_{\rm c}^*$; no roughening occurs. 
\item For $\xi/L < C_{\rm V}^2$ (strong disorder): $\lambda_{\rm c} < l < \lambda_{\rm c}^*$; the crack roughens on scales between $\lambda_{\rm c}$ and $\lambda_{\rm c}^*$.
\end{itemize}
With roughness exponent $\zeta$ and upper correlation length $\lambda_{\rm c}^*$, we expect the characteristic width of a depinning crack to be given by
\begin{equation}
(\langle \delta l^2\rangle)^{1/2} \approx \xi \left(\frac{\lambda_{\rm c}^*}{\lambda_{\rm c}}\right)^\zeta \approx \xi \left(\frac{l}{\xi}\right)^{2\zeta} C_{\rm V}^{4 \zeta}\;.
\label{cracwidth}
\end{equation}

\section{Comparison with simulation results}

To test the scaling arguments developed in the previous section, we simulate the evolution of cracks under conditions of load control. We use a lattice automaton technique where we evaluate the displacements at discrete sites $(x_i,y_j)$ on a two-dimensional lattice with lattice constant $\xi$. Instead of brittle behavior ($\tau(u) = W_{\rm f} \delta(u)$) we consider a triangle-shaped stress-displacement characteristics 
\begin{equation}
\tau(u) = \left\{\begin{array}{ll}
\tau_{\rm M} \displaystyle\frac{u}{u_0} \;,& u<u_0\;,\\
\tau_{\rm M} \displaystyle(2-\frac{u}{u_0}) \;,& u_0 \le u < 2 u_0\;,\\
0\;,& u\ge 2u_0\;.\end{array}
\right.
\label{semibrittle}
\end{equation}
The fracture energy is here related to the peak strength $\tau_{\rm M}$ and characteristic displacement-to-fracture $u_0$ by $W_{\rm f} = \tau_{\rm M} u_0$. Disorder is introduced by considering the peak strengths (or equivalently the fracture energies) at different sites as independent Weibull distributed random variables. In the following we use nondimensional variables defined through
\begin{equation}
T=\frac{\tau }{\left\langle \tau _{{\rm M}} \right\rangle } \; ,\; \Sigma = \frac{\sigma _{xz}^{{\rm EXT}} }{\left\langle \tau _{{\rm M}} \right\rangle } \;,\; U=\frac{u}{u_{0} } \; ,\; X=\frac{x}{\xi} \;, Y=\frac{y}{\xi} \;,
\label{scale}
\end{equation}
such that the scaled average peak strength $\langle T \rangle$ and fracture energy $\langle W \rangle$ are by definition equal to 1. The scaled crack width is accordingly defined as $L = l/\xi$. Furthermore, we make the simplifying assumption that the mode-II and mode-III interaction constants are equal, $I_x = I_y = I$. The equilibrium equation then reads
\begin{equation}
J \left(\frac{\partial ^{2} U}{\partial X^{2} } + \frac{\partial ^{2} U}{\partial Y^{2} }\right) + \Sigma - T \le 0\;,
\label{equil_U}
\end{equation}
where the non-dimensional interaction constant is $J = (I u_0)/(\tau_{\rm M} \xi^2)$. The non-dimensional crack driving force (stress intensity) is given by $F = (L\Sigma)^2/(2 J)$ and accordingly the length of the critical reference crack in a homogeneous system is obtained from the requirement $F_{\rm c} = 1$ as $L_{\rm c} = \sqrt{2 J}/\Sigma$. In the simulations we evaluate $U_{XX}$ and $U_{YY}$ in terms of the corresponding discrete second order gradients.

\begin{figure}[tbh]
\begin{center}
\includegraphics{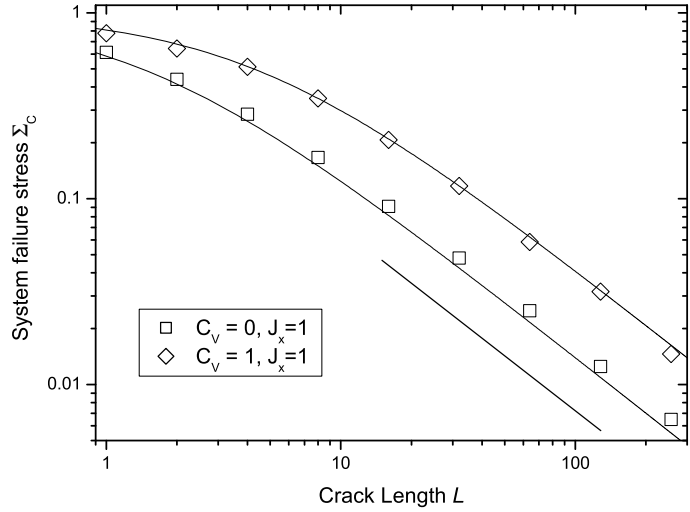}
\end{center}
\caption{System failure stress versus crack length for systems without fluctuations and two different values of the interaction parameter $J$; full lines: analytical approximations according to Eqn. (\ref{eq:size}). }
\label{sizeffect}
\end{figure}

Simulation of a system containing a shear crack is carried out as follows: To create a mode-II crack of non-dimensional length $2 L$, the strength is set to zero over a `strip' of width $2 L$ and $U$ is assumed everywhere zero. Then, the system is loaded by increasing the external stress $\Sigma$ from zero in small steps $\Delta \Sigma$. When sites become unstable as the local (external plus internal) stress exceeds the local strength, the displacement at all unstable sites is increased by a small amount $\Delta U$. Then, new internal stresses are re-computed and it is checked again where the sum of the external and internal stresses exceeds the local strength. The displacement at the now unstable sites is again increased, etc. This is repeated until the system has reached a new stable configuration. Then the external stress is increased again and the procedure is repeated until the system fails completely ($U_{i} > 2$ for all sites). The corresponding \textit{critical stress} is denoted by $\Sigma_{C}$. The procedure is repeated for different values of $\Delta U$ and $\Delta \Sigma$ to ensure that the results do not depend on step size. The stress-dependent evolution of the crack front (defined as the boundary of the area where $U_i > 2$) is recorded, as is the failure stress.

To illustrate how the simulations, which assume semi-brittle behavior (finite values of $\tau_{\rm M}$ and $u_0$), compare with the analytical results for the brittle limit, we show in Figure 1 the dependence of failure stress on crack length $L$ for two systems without disorder. The failure stresses are well described by the analytical equation
\begin{equation}
\Sigma_{\rm c} = \frac{\sqrt{2J}}{L + \sqrt{2J}}\;,
\label{eq:size}
\end{equation}
which for $L \gg J$ approaches the result for the brittle case, $\Sigma_{\rm c} = \sqrt{2J}/L$. The analytical form of Eqn. (\ref{eq:size}) is in line with the suggestion of Bazant \cite{bazant98} to introduce size-dependent corrections into Griffith-like criteria for semibrittle materials by replacing the crack length with the crack length plus process zone size. In fact, it is easy to see that $\sqrt{J}$ defines the non-dimensional characteristic length of the 'process zone' ahead of the crack tip which derives from solving Eq. (\ref{equil_U}) for the semibrittle law given by Eqn. (\ref{semibrittle}). In comparing simulations and scaling results we shall in the following exclusively consider situations close to the brittle limit, $L > 20 J$, where corrections due to finite process zone size are small.

\begin{figure}[tbh]
\begin{center}
\includegraphics{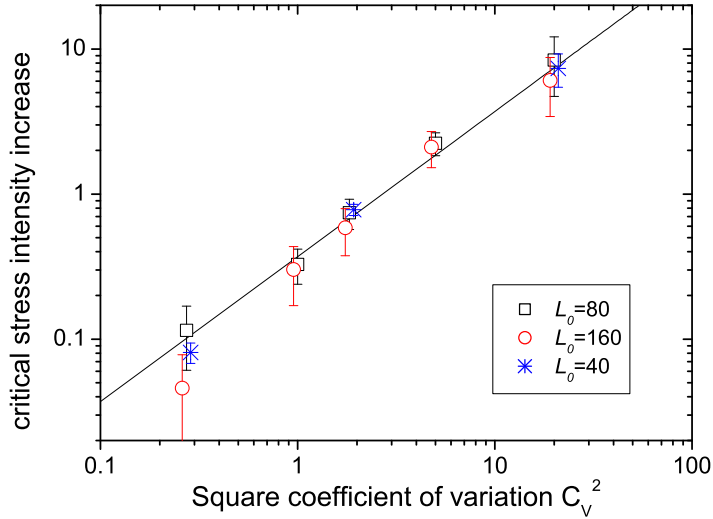}
\end{center}
\caption{Crack pinning: Increase in critical stress intensity with increasing disorder; initial crack length $L_0$=40, $L_0$=80,
$L_0$=160, interaction parameter $J=1$; full line: $\Delta F_{\rm c} \propto C_{\rm V}^2$.}
\label{pinning}
\end{figure}

\subsection{Crack pinning}

We now turn to systems with disorder and investigate the disorder-induced increase in the critical force required for unstable crack propagation. To this end, we adjust the parameters of the Weibull distribution of $T$ such that the average of the distribution is kept at $\langle T \rangle = 1$ while the coefficient of variation, which serves as a measure of the degree of disorder, is varied over a series of simulations. Again we introduce cracks of different initial length and increase the external stress on the system until failure occurs. This is in general preceded by subcritical crack advances between intermediate pinned configurations. We record the mean crack length $L = \langle L(Y) \rangle$ together with the stress such that we can determine the non-dimensional crack driving force (stress intensity) as $F = (\langle L(Y) \rangle \Sigma)^2/(2 J)$. The disorder-induced increase in critical stress intensity at depinning, $\Delta F_{\rm c} = F_{\rm c} - 1$, is shown in Figure \ref{pinning} as a function of the coefficient of variation of the fracture energy distribution. It is seen that the simulation results are in good agreement with Eq. (\ref{cracpin2}) which predicts a quadratic relationship between the increase in critical stress intensity $F_{\rm c}$ and the coefficient of variation of the fracture energy distribution. Also in agreement with Eq. (\ref{cracpin2}), it is found that the increase in critical stress intensity does not appreciably depend on the length of the crack. 

\subsection{Crack roughening and final crack shape}

\begin{figure}[tbh]
\begin{center}
\includegraphics{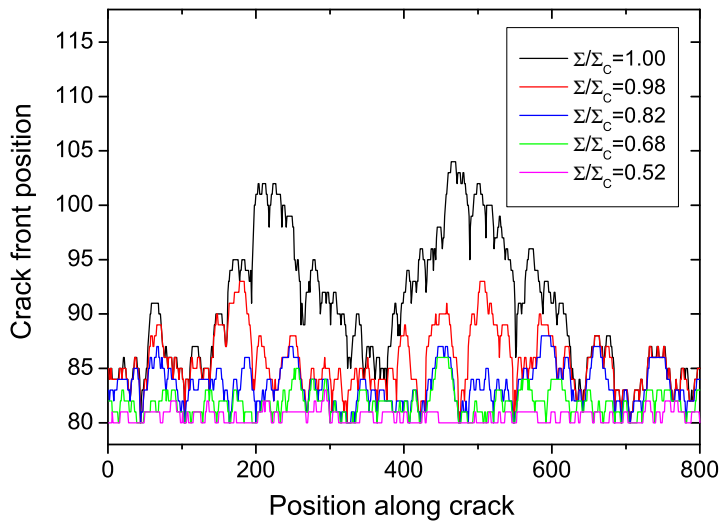}
\end{center}
\caption{Sequence of crack front positions at increasing external stress; initial crack length $L_0$=80, interaction parameter $J=1$, Weibull modulus $\beta = 0.5$. The uppermost curve gives the crack shape immediately before system failure by unstable crack propagation.}
\label{crackshapes}
\end{figure}

We now study how the crack shape evolves during loading. Simulations were carried out for cracks of varying initial length $L_0$ in the regime $40 < L_0/J < 160$ where according to Figure 1 the brittle approximation can be considered valid. Figure \ref{crackshapes} shows a crack shape sequence that occurs while the external load is increased until failure. The crack develops a complex shape which we analyze in terms of self-affine roughness. To demonstrate self-affine roughness we perform a multiscaling analysis where we evaluate the $n$-th order structure function $C_n(S) := \langle |L(Y)-L(Y+S)|^n|\rangle^{1/n}$ (the average runs over all positions $Y$ in a given realization and over multiple realizations of the disorder). Self-affine behavior is characterized by a power-law relationship $C_n(S) \propto S^{\zeta_n}$
where $\zeta_n = \zeta$ does not depend on the order of the structure function. 

\begin{figure}[tbh]
\begin{center}
\includegraphics{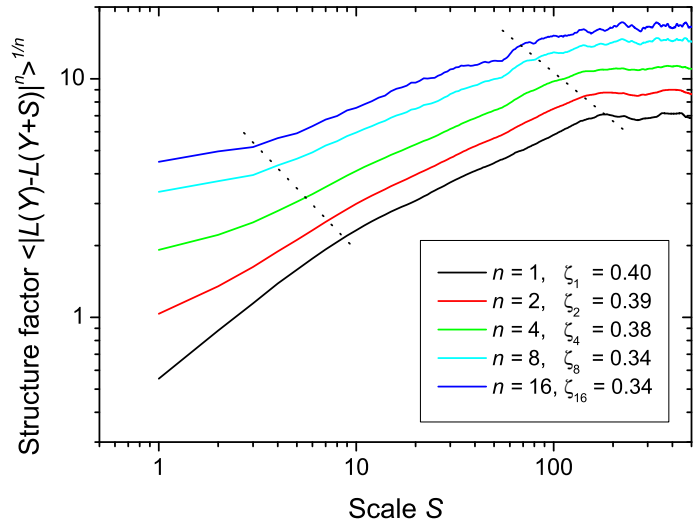}
\end{center}
\caption{Multiscaling analysis of crack shapes immediately before failure; parameters as in Figure \ref{crackshapes}; the dotted lines indicate the limits of the scaling regimes used in fitting the scaling exponents $\zeta_n$.}
\label{multiscaling}
\end{figure}

Figure \ref{multiscaling} shows that power-law scaling is indeed observed over about 1.5 orders of magnitude in scale. Within the scaling regimes and for $1 \le n \le 16$, we find only a very weak n-dependence of the scaling exponent $\zeta_n = 0.35\dots 0.4$, justifying the conjecture of self-affine behavior. The $\zeta$ exponent values of about 0.4 are in approximate agreement with the theoretical results and experimental observations quoted in the introduction. It is important to note that the scaling regimes in our simulations are quite limited: Self-affine scaling is limited from below by the pinning length $L_{\rm c}$ and from above by the intrinsic correlation length $L_{\rm c}^* = \lambda_{\rm c}^*/\xi$. 

\begin{figure}[tbh]
\begin{center}
\includegraphics{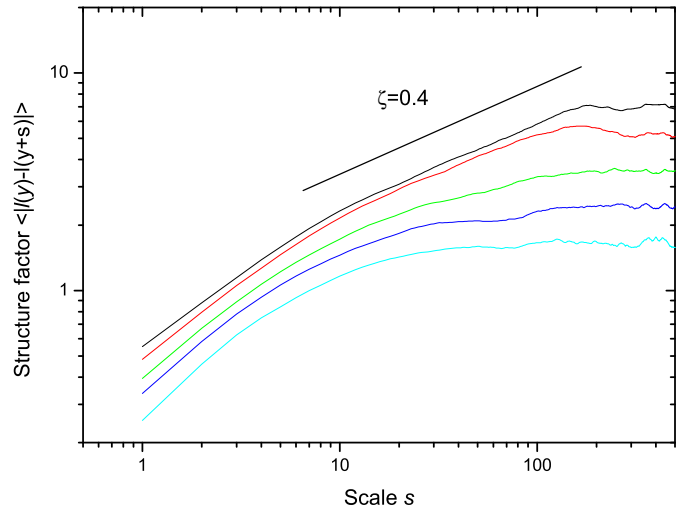}
\end{center}
\caption{Evolution of the structure factor with increasing load. Values of the excess driving force from bottom to top:
$\Delta F/\Delta F_{\rm c} = 0.2, 0.4, 0.6, 0.8, 1.0$.}
\label{roughening}
\end{figure}

An analysis of the roughening kinetics has been performed in terms of the lowest-order structure factor $C_1(S)$. This is shown in Figure \ref{roughening} where $C_1(S)$ is plotted for various levels of the crack driving force. As the force increases
towards the depinning threshold, the crack roughness increases on all scales and a linear scaling regime emerges in the 
double-logarithmic plot. However, it is evident from the figure that the roughening kinetics cannot be described by the usual Family-Vicsek form \cite{family85}.

\begin{figure}[tbh]
\begin{center}
\includegraphics{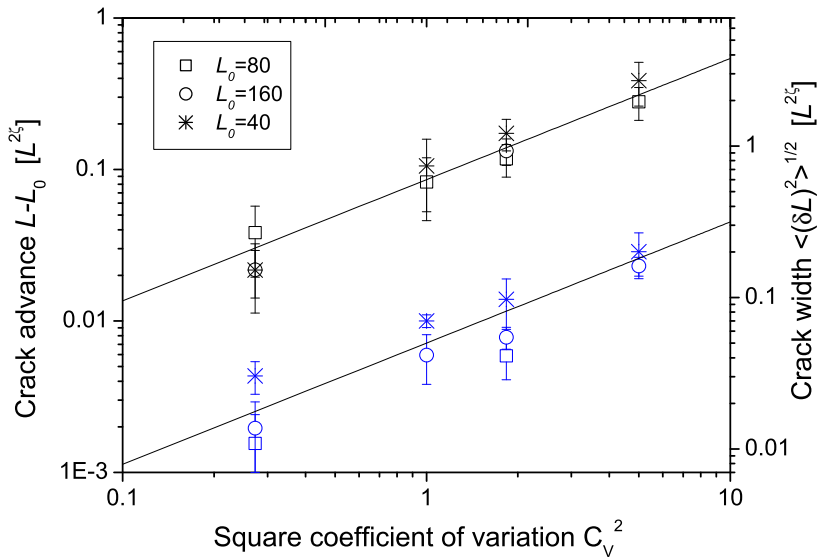}
\end{center}
\caption{Width of the critical crack (lower symbols, right axis) and crack length increase $\langle L \rangle - L_0$ (upper symbols, left axis) as a function of the coefficient of variation of the fracture energy distribution; all lengths have been 
scaled with $L^{2\zeta}$ where $\zeta = 0.4$; the full lines have slope $2 \zeta = 0.8$; note that the left and right axes have been shifted for improved visibility. The error bars represent the variance of values determined from 10 simulations}
\label{crackwidth}
\end{figure}

In spite of this complication, the final shape of the critical crack follows the results of the scaling theory outlined in 
Section 3. Figure \ref{crackwidth} shows how the width of critical cracks depends on the degree of disorder. In line with Eqn. (\ref{cracwidth}), the width of the crack front scales like $L^{2 \zeta} C_{\rm V}^{4 \zeta}$ where $\zeta \approx 0.4$. The same scaling applies to the difference $L - L_0$ of the initial and final crack front positions. This is to be expected 
since the crack roughens by advancing. 

\section{Discussion and Conclusions}

\begin{figure}[tbh]
\begin{center}
\includegraphics{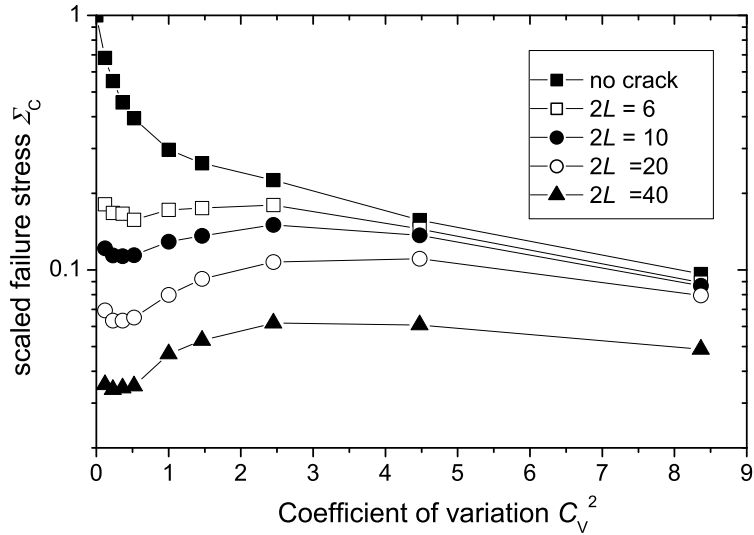}
\end{center}
\caption{Critical stress for system failure as a function of the coefficient of variation of the fracture energy distribution for different initial crack lengths} 
\label{failurestress}
\end{figure}

We have shown that disorder leads to crack front roughening while simultaneously increasing the fracture toughness of the material. Even though the critical stress intensity required for unstable crack propagation increases with increasing disorder, under conditions of load control this does {\em not} necessarily mean that the stress supported by the system (with or without crack) increases as well. Figure \ref{failurestress} shows instead a complicated behavior: The strength of cracked interfaces first decreases with increasing disorder, then increases, then again decreases. 
To understand this complex behavior we have to remember that the presence of disorder allows the crack to advance by some amount $\Delta L$ even before it becomes critical. This subcritical advance increases the driving force (stress intensity factor) acting on the crack front. At the same time, however, the critical force required for crack propagation increases as well by an amount $\Delta F$. Using the condition for the crack to become critical, $(\Sigma L)^2/(2J) = (1+\Delta F)$, it is easy to see that the resulting change $\Delta \Sigma$ in critical stress fulfills the approximate relation
\begin{equation}
\frac{\Delta \Sigma}{\Sigma} \approx \frac{\Delta F}{2} - \frac{\Delta L}{L}\;.
\end{equation}
Since $\Delta L \propto C_{\rm V}^{4\zeta} \approx C_{\rm V}^{1.6}$ while $\Delta F \propto C_{\rm V}^{2}$, the softening effect of the increasing crack length prevails at small disorder, while the strengthening effect of crack front pinning takes over at larger disorder, giving rise to the observed non-monotonic behavior. 

However, why does the strength decrease again at very large disorder? This effect cannot be understood by considering exclusively the dynamics of the pre-existing crack. Rather, one has to allow for the nucleation of new cracks at other sites within the system which is facilitated by the disorder as shown in \cite{zaiser09}. In the regime of large disorder, propagation of the existing crack is inhibited while nucleation of new cracks at weak sites elsewhere in the system is facilitated. As a consequence, one ultimately reaches a situation where failure occurs not by propagation of the pre-existing crack (which instead becomes arrested by the disorder) but by nucleation of one or several cracks at other, more favorable sites. 

In conclusion, we have demonstrated that the theory of elastic manifold depinning offers a useful framework for understanding the behavior of cracks in disordered interfaces underneath thin films. Standard scaling arguments can be used to predict the influence of disorder on the critical stress intensity factor required for sustained crack propagation, and the geometry of the depinning crack front exhibits features characteristic of a depinning manifold with long-range elastic interactions. However, at the same time the standard depinning framework needs to be modified in several respects: (i) On large scales, the finite crack length $l$ leads to a change in the elastic interactions and to the emergence of an intrinsic correlation length 
$\lambda_{\rm c}^* > l$  above which the crack remains flat; (ii) The driving force acting on the crack (the stress intensity factor) depends on the crack length and can therefore not be controlled independently. Subcritical crack roughening proceeds through advance of the crack front which increases the stress intensity. As a consequence, one finds a complicated disorder dependence of the system failure stress. (iii) At very large disorder, the failure mode changes and crack nucleation outwith  pre-existing cracks becomes predominant. The system strength in this regime is expected to decrease with increasing system size and disorder, similar to the behavior of one-dimensional systems studied in \cite{zaiser09}. A study of crack nucleation in two-dimensional systems remains a task for future investigations.

\ack
Financial support of the European Commission under contract
NEST-2005-PATH-COM-043386 (NEST pathfinder programme TRIGS)  
is gratefully acknowledged.


\end{document}